\documentclass[aps,prl,twocolumn,groupedaddress]{revtex4-1}
\usepackage{graphicx}

\begin{document}


\title{Probing the Superfluid to Mott Insulator Transition at the Single Atom
Level}


\author{Waseem S. Bakr}
\author{Amy Peng}
\author{M. Eric Tai}
\author{Ruichao Ma}
\author{Jonathan Simon}
\author{Jonathon I. Gillen}
\author{Simon F\"olling}
\altaffiliation{Institut f\"ur Physik, Ludwig-Maximilians-Universit\"at, 80799 M\"unchen, Germany}
\author{Lode Pollet}
\author{Markus Greiner}
\email[]{greiner@physics.harvard.edu}
\affiliation{Department of Physics, Harvard University, Cambridge,
Massachusetts, 02138, USA}


\date{\today}

\begin{abstract}
Quantum gases in optical lattices offer an opportunity to experimentally
realize and explore condensed matter models in a clean, tunable system. We
investigate the Bose-Hubbard model on a microscopic level using single
atom--single lattice site imaging; our technique enables space- and
time-resolved characterization of the number statistics across the
superfluid--Mott insulator quantum phase transition. Site-resolved probing of
fluctuations provides us with a sensitive local thermometer, allows us to
identify microscopic heterostructures of low entropy Mott domains, and enables
us to measure local quantum dynamics, revealing surprisingly fast transition
timescales. Our results may serve as a benchmark for theoretical studies of
quantum dynamics, and may guide the engineering of low entropy phases in a
lattice.
\end{abstract}


\maketitle

Microscopic measurements can reveal properties of complex systems that are not
accessible through statistical ensemble measurements. For example, scanning
tunneling microscopy has allowed physicists to identify the importance of
nanoscale spatial inhomogeneities in high temperature superconductivity
\cite{lang_imaginggranular_2002}, and single molecule microscopy
\cite{moerner_single-molecule_1996} has enabled studies of local dynamics in
chemical reactions revealing e.g. the importance of multiple reaction pathways
\cite{zhuang_single-molecule_2000}.  While previous ultracold quantum gas
experiments have focused primarily on statistical ensemble measurements, the
recently introduced single atom–-single lattice site imaging technique in a
Quantum Gas Microscope (QGM) \cite{bakr_quantum_2009} opens the door for
probing and controlling quantum gases on a microscopic level. Here we present
a microscopic study of an atom-lattice system that realizes the bosonic
Hubbard model and exhibits a quantum phase transition from a superfluid to a
Mott insulator \cite{fisher_boson_1989,jaksch_cold_1998,greiner_quantum_2002}.
In the weakly interacting superfluid regime, the many-body wavefunction
factorizes into a product of states with well-defined phase on each lattice
site, known as coherent states, with Poissonian number fluctuations. As the
strength of the interaction increases, the number distribution is narrowed,
resulting in a fixed atom number state on each site deep in the Mott insulator
regime. We study this change in the number statistics across the transition;
these microscopic studies are complementary to previous experiments that have
focused on measuring ensemble properties such as long range phase coherence,
excitation spectra or compressibility
\cite{greiner_quantum_2002,jordens_mott_2008,schneider_metallic_2008}. Local
properties such as onsite number statistics
\cite{capogrosso-sansone_-site_2007} were accessible only indirectly
\cite{greiner_collapse_2002,gerbier_probing_2006,jordens_mott_2008} and
averaged over several shells of superfluid and Mott insulating domains in the
inhomogeneous system, complicating quantitative interpretation. More recently,
the shell structure was imaged through tomographic
\cite{foelling_formation_2006}, spectroscopic
\cite{campbell_imagingmott_2006}, and in-situ imaging techniques,
coarse-grained over several lattice sites \cite{gemelke_in_2009}.

\begin{figure}
    \centering
    \includegraphics[width=8.5cm]{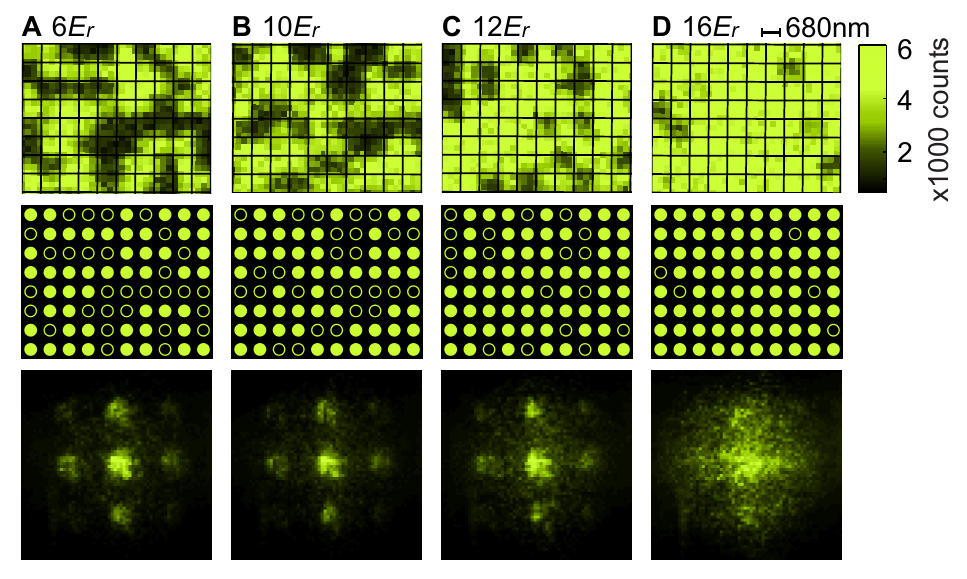}
    \caption{\label{fig1} Single-site imaging of atom number fluctuations
    across the superfluid-Mott insulator transition. ({\bf A} -- {\bf D})
    Images within each column are taken at the same final 2D lattice depth of
    ({\bf A}) $6E_r$, ({\bf B}) $10E_r$, ({\bf C}) $12E_r$ and ({\bf D})
    $16E_r$. Top row: in-situ fluorescence images from a region of $10\times8$
    lattice sites within the $n=1$ Mott shell that forms in a deep lattice. In
    the superfluid regime $({\bf A}, {\bf B})$, sites can be occupied with odd
    or even atom numbers, which appear as full or empty sites respectively in
    the images. In the Mott insulator, occupancies other than 1 are highly
    suppressed ({\bf D}). Middle row: results of the atom detection algorithm
    \cite{som}  for images in the top row. A full (empty) circle indicates the
    presence (absence) of an atom on a site. Bottom row: time of flight
    fluorescence images after 8ms expansion of the cloud in the 2D plane as a
    result of non-adiabatically turning off the lattice and the transverse
    confinement (averaged over 5 shots and binned over $5\times5$ lattice
    sites).}
\end{figure}

We start with a two-dimensional $^{87}$Rb Bose-Einstein condensate of a few
thousand atoms confined in a single well of a standing wave, with a harmonic
oscillator length of 130nm\cite{som}. The condensate resides $9\mu$m from an
in-vacuum lens that is part of an imaging system with a resolution of
$\sim600$nm. This high resolution system is used to project a square lattice
potential onto the pancake cloud with a periodicity of $a=680$nm, as described
in previous work \cite{bakr_quantum_2009}. The lattice depth is ramped
exponentially with a time constant of 81ms up to a maximum depth of $16E_r$,
where $E_r$ is the recoil energy of the effective lattice wavelength given by
$h^2/8ma^2$, with $m$ being the mass of $^{87}$Rb and Planck's constant $h$.
In a homogeneous system in two dimensions, the transition to a Mott insulator
with one atom per site occurs at a ratio of interaction energy to tunneling of
$U/J=16.7$
\cite{kohl_superfluid_2005,spielman_mott-insulator_2007,capogrosso-sansone_monte_2008},
corresponding to a lattice depth of $12.2E_r$. During this ramp, the initial
transverse confinement of 9.5Hz is increased such that the cloud size remains
approximately constant. After preparing the many-body state, we image the
atoms by increasing the lattice depth several hundred-fold, and then
illuminate the atoms with an optical molasses that serves to localize the
atoms while fluorescence photons are collected by the high resolution optics.
As a result of the imaging process, the many-body wavefunction is projected
onto number states on each lattice site. In addition, light-assisted
collisions immediately eject atoms in pairs from each lattice site, leaving
behind an atom on a site only if its initial occupation was odd
\cite{depue_unity_1999}. Remaining atoms scatter several thousand photons
during the exposure time and can be detected with high fidelity. By preparing
the sample repeatedly under the same conditions, we deduce the probability
$p_{\mbox{\scriptsize \emph{odd}}}$ of having an odd number of atoms on a site
before the measurement.

For a coherent state on a lattice site with mean atom number $\lambda$,
$p_{\mbox{\scriptsize \emph{odd}}}$ is given by $1/2 (1-e^{-2\lambda})< 1/2$.
In a Mott-insulating region in the zero temperature and zero tunneling limit,
$p_{\mbox{\scriptsize \emph{odd}}}$ is 1 (0) for shells with an odd (even)
atom number per site. Fig. \ref{fig1} shows fluorescence images in a region of
the cloud as the final depth of the lattice is increased. The initial
superfluid density is chosen to obtain an insulator with two shells on the
Mott side of the transition, and the region shown is in the outer shell
containing one atom per site. For high filling fractions, the lattice sites in
the images are barely resolved, but the known geometry of the lattice and
imaging system point spread function obtained from images at sparser fillings
allow reliable extraction of site occupations \cite{som}.

\begin{figure}
    \centering
    \includegraphics[width=8.5cm]{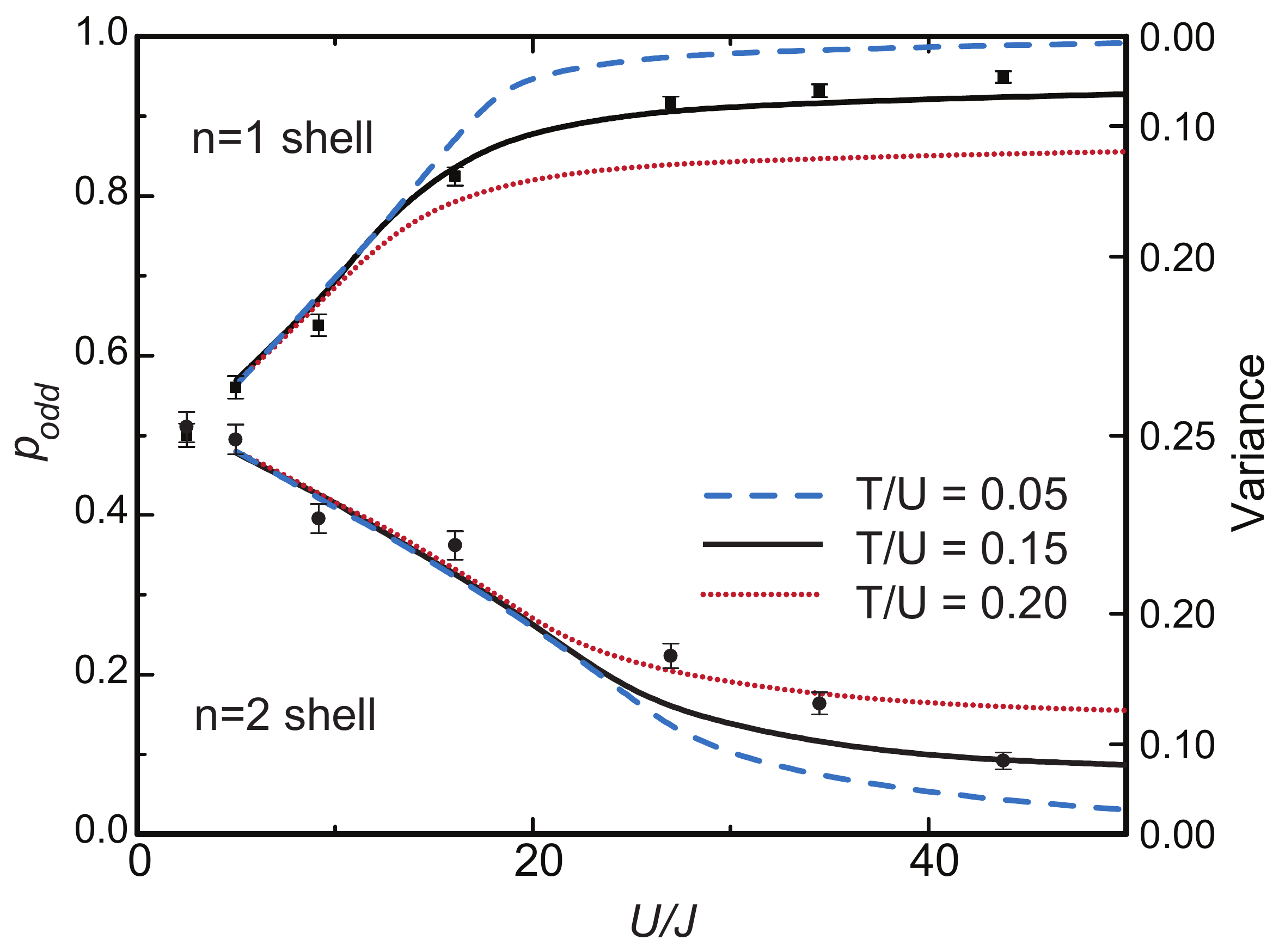}
    \caption{\label{fig1E} Measured value of $p_{\mbox{\scriptsize
    \emph{odd}}}$ vs. the interaction to tunneling ratio ($U/J$). Data sets,
    with statistical error bars, are shown for regions that form part of the
    $n=1$ (squares) and $n=2$ (circles) Mott shells in a deep lattice. The
    lines are based on finite temperature Monte-Carlo simulations in a
    homogeneous system at constant temperature to interaction ratio ($T/U$) of
    0.20 (dotted red), 0.15 (solid black) and 0.05 (dashed blue). The axis on
    the right is the corresponding odd-even variance given by
    $p_{\mbox{\scriptsize \emph{odd}}}(1- p_{\mbox{\scriptsize
    \emph{odd}}})$.}
\end{figure}

We determine $p_{\mbox{\scriptsize \emph{odd}}}$ for each site using 24 images
at each final lattice depth. The transverse confining potential varies slowly
compared to the lattice spacing and the system is to good approximation
locally homogeneous. We make use of this to improve the error in our
determination of $p_{\mbox{\scriptsize \emph{odd}}}$, by averaging over a
group of lattice sites, in this case 51 (30) sites for regions in the first
(second) shell (Fig. \ref{fig1E}). In the $n=1$ shell, we
detect an atom on a site with probability $94.9\pm 0.7\%$ at a lattice depth of $16 E_r$. We
measure the lifetime of the gas in the imaging lattice and determine that
$1.75\pm0.02\%$ of the occupied sites are detected as unoccupied due to atoms lost
during the imaging exposure time of 1s because of background gas collisions.
We correct for this effect in the given average occupation numbers and errorbars.

Measuring the defect density in the Mott insulator provides sensitive local
thermometry deep in the Mott regime. Thermometry in the Mott state has been a
long-standing experimental challenge
\cite{weld_spin_2009,trotzky_suppression_2009} and has acquired particular
significance as experiments approach the regime of quantum magnetism
\cite{duan_controlling_2003,altman_phase_2003,trotzky_time-resolved_2008}
where the temperature scale should be on the order of the superexchange
interaction energy. We directly image excitations of the $n=1$ Mott insulator,
holes and doublons, as they both appear as missing atoms in the images.
Similarly, for Mott insulators with higher fillings $n$, sites with
excitations $(n+1, n-1)$ can be detected through their opposite parity signal.
For finite tunneling rate $J$ much smaller than the interaction energy $U$,
the admixture fraction of coherent hole-doublon pairs excitations is
$\sim(J/U)^2$, whereas any other excitations are due to incoherent thermal
fluctuations and are suppressed by a Boltzmann factor $e^{-U/T}$.

The theory curves presented in Fig. \ref{fig1E} are the predicted $p_{\mbox{\scriptsize
\emph{odd}}}$ in the two shells for different values of $T/U$. The curves are
obtained using a quantum Monte-Carlo ``worm'' algorithm
\cite{prokofev_worm_1998,pollet_engineering_2007}, and the average temperature
extracted using the data points at the three highest $U/J$ ratios is
$T/U\sim0.16 \pm 0.03$. At the transition point for $n=1$, this corresponds to
a temperature of 1.8nK. Assuming this value of $T/U$ to be the overall
temperature, the thin layer between the Mott shells should be superfluid, and
the transition to a normal gas is expected around a critical temperature of
$zJ=2.8$nK, where $z$ is the number of nearest neighbors in the lattice
\cite{gerbier_boson_2007}.

\begin{figure}
    \includegraphics[width=8.5cm]{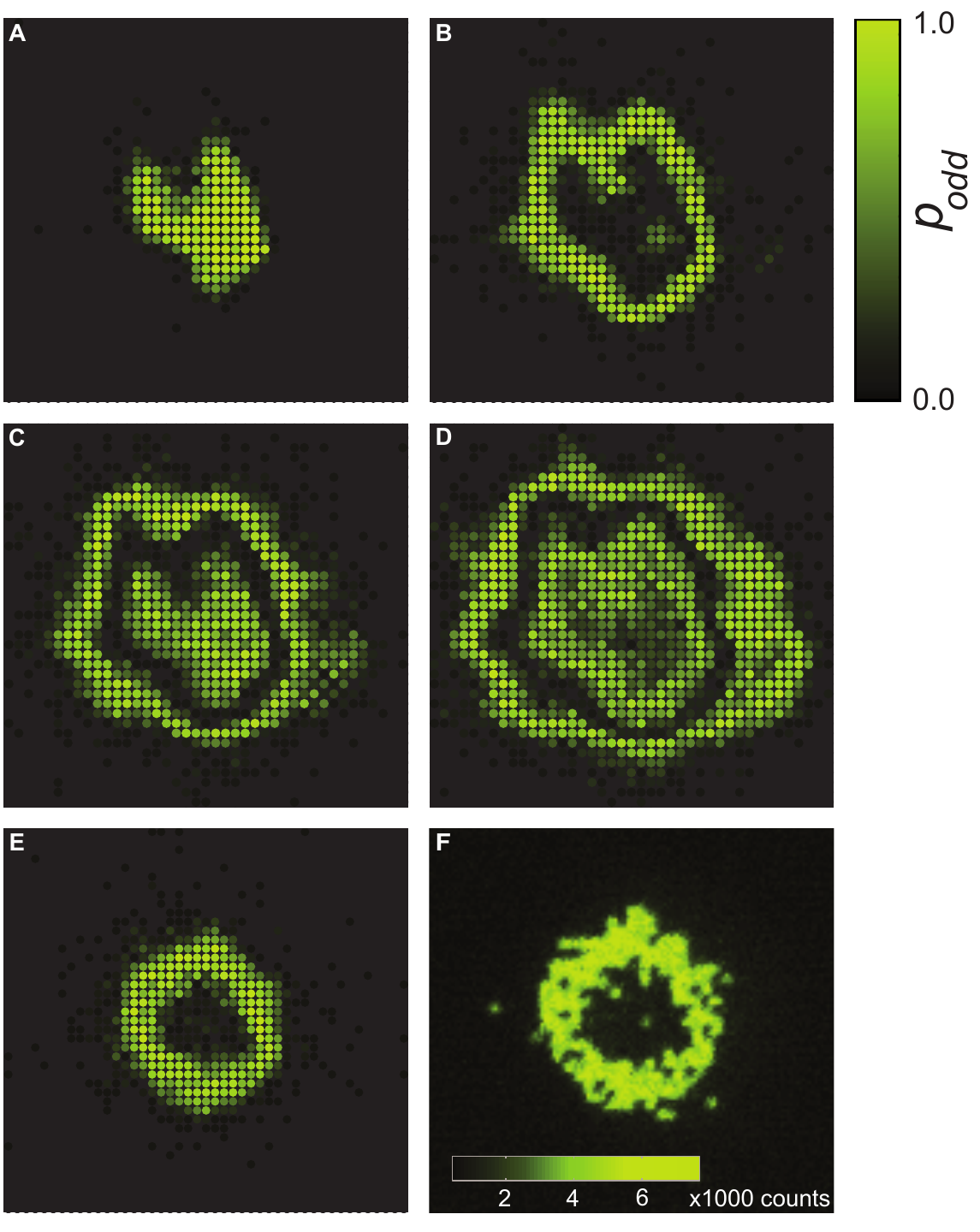}
    \caption{\label{fig2} Single-site imaging of the shell structure in a Mott
    insulator. ({\bf A} -- {\bf D}) The images show $p_{\mbox{\scriptsize
    \emph{odd}}}$ on each site determined by averaging 20 analyzed
    fluorescence images. The lattice depth is $22E_r$ and the transverse
    confinement is 45Hz. As the atom number is increased from ({\bf A}) to
    ({\bf D}), the number of shells in the insulator increases from one to
    four. The value of $p_{\mbox{\scriptsize \emph{odd}}}$ for odd (even)
    numbered shells is close to one (zero). The atom numbers, determined by
    in-situ imaging of clouds expanded in the plane, are ({\bf A}),
    $120\pm10$, ({\bf B}), $460\pm20$, ({\bf C}), $870\pm40$ and ({\bf D}),
    $1350\pm70$. ({\bf E}--{\bf F}) Long wavelength disorder can be corrected
    by projecting an appropriate compensation light pattern onto the atoms,
    resulting in nearly circular shells. ({\bf E}) shows $p_{\mbox{\scriptsize
    \emph{odd}}}$ (average of 20 analyzed images) and ({\bf F}) is a single
    shot raw image (arbitrary units).}
\end{figure}

Next we study the global structure of the Mott insulator.  The high resolution
images provide an atom-by-atom picture of the concentric shell structure,
including the transition layers in between the insulating shells. In Fig.
\ref{fig2}A to D, the formation of the various shells, up to the fourth, is
shown as the atom number in the trap is increased. Slowly varying optical
potential disorder causes deviation from circular symmetry in the shells. The
contour lines of the potential are directly seen in the images in Fig.
\ref{fig2}. In Fig. \ref{fig2}E and F, we have compensated this disorder by
projecting a light pattern generated using a digital micromirror device
through the objective\cite{som}, resulting in a nearly circular shell
structure.

\begin{figure}
    \centering
    \includegraphics[width=8.5cm]{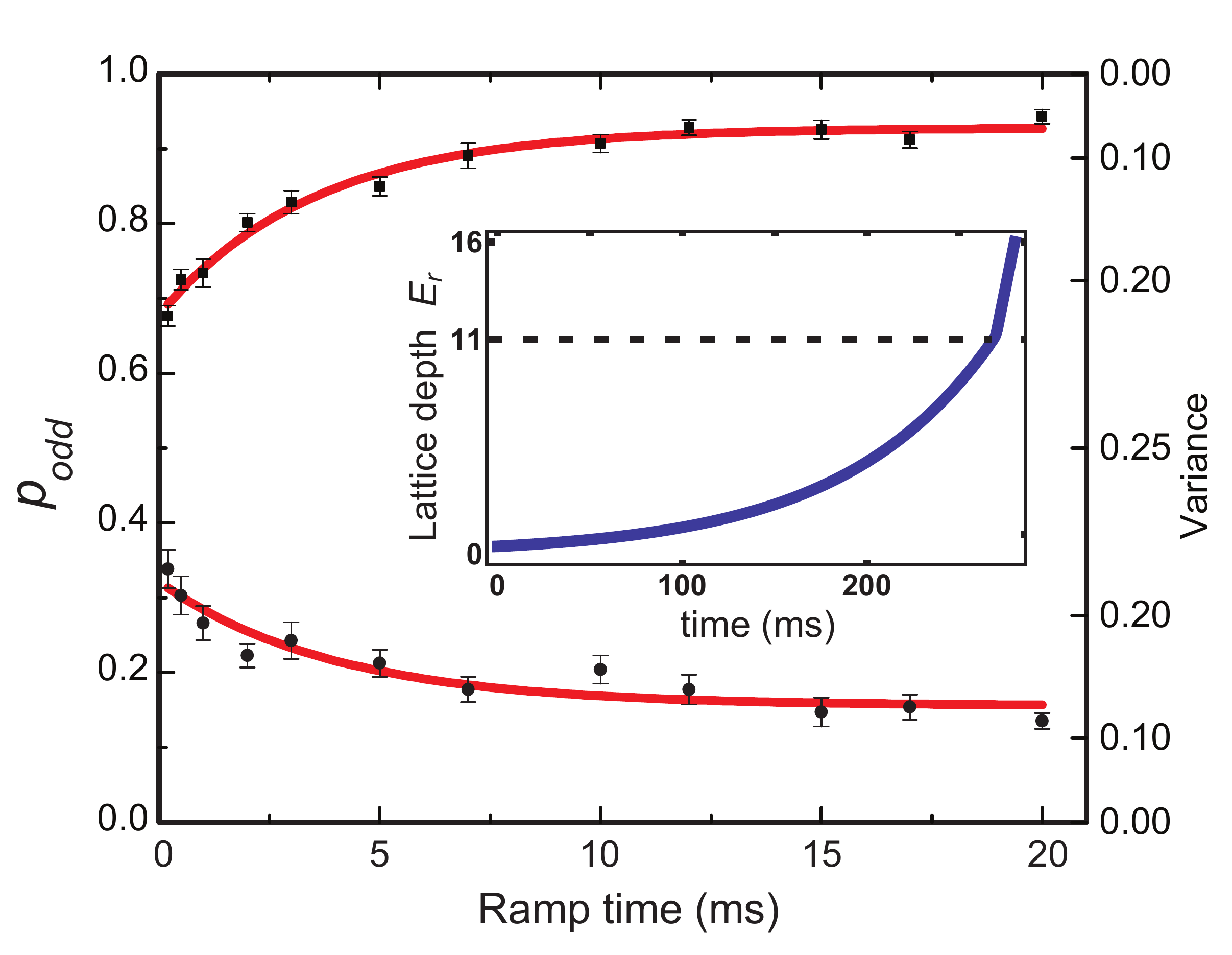}
    \caption{\label{fig3} Dynamics of on-site number statistics for a fast
    ramp from the superfluid regime to the Mott regime. $p_{\mbox{\scriptsize
    \emph{odd}}}$ at the end of the ramp vs. ramp time is shown in the $n=1$
    (squares) and $n=2$ (circles) shells, averaged over 19 datasets with
    statistical errorbars. Red lines are exponential fits. Inset: the two-part
    ramp used in this experiment. The first part is a fixed adiabatic
    exponential ramp ($t=81$ms) and the second is a linear ramp starting at
    $11E_r$ and ending at $16E_r$. The duration of the second ramp is varied
    in the experiment.}
\end{figure}

In a second series of experiments, we use on-site number statistics to probe
the adiabaticity timescale for the transition, focusing on the local dynamics
responsible for narrowing the number distribution. We start by increasing the
lattice depth adiabatically to $11E_r$, still in the superfluid regime, using
the same ramp described previously. Next the depth is ramped linearly to
$16E_r$ where, for an adiabatic ramp, a Mott insulator should form. The ramp
time is varied from 0.2ms to 20ms, and $p_{\mbox{\scriptsize \emph{odd}}}$ is
measured in the first and second shells as before (Fig. \ref{fig3}); we find
that the data fits well to exponential curves that asymptote to the value of
$p_{\mbox{\scriptsize \emph{odd}}}$ obtained in the adiabatic case. The fitted
time constant in the first (second) shell is $3.5\pm0.5$ms ($3.9\pm1.3$ms).

Compared to the critical value of the tunneling time $h/J_c=68$ms for the first
shell, the observed dynamics are counter-intuitively fast. This can be
understood using a simple picture of two atoms in a double well. In this
system, as the tunneling is varied, the minimal gap between the ground state
and the first excited state is $U$, which sets the adiabaticity timescale. It
is an open question whether this argument can be generalized to a lattice. In an
infinite system, the appearance of Goldstone modes in the superfluid regime
leads to a vanishing gap at the transition point, but the density of states is
low for energies much less than $U$ \cite{knap_spectral_2010}. In fact, the
$1/e$ timescale observed experimentally is comparable to
$h/U_c=4.1$ms, where $U_c$ is the critical interaction energy for an $n=1$
insulator.

\begin{figure}
    \centering
    \includegraphics[width=8.5cm]{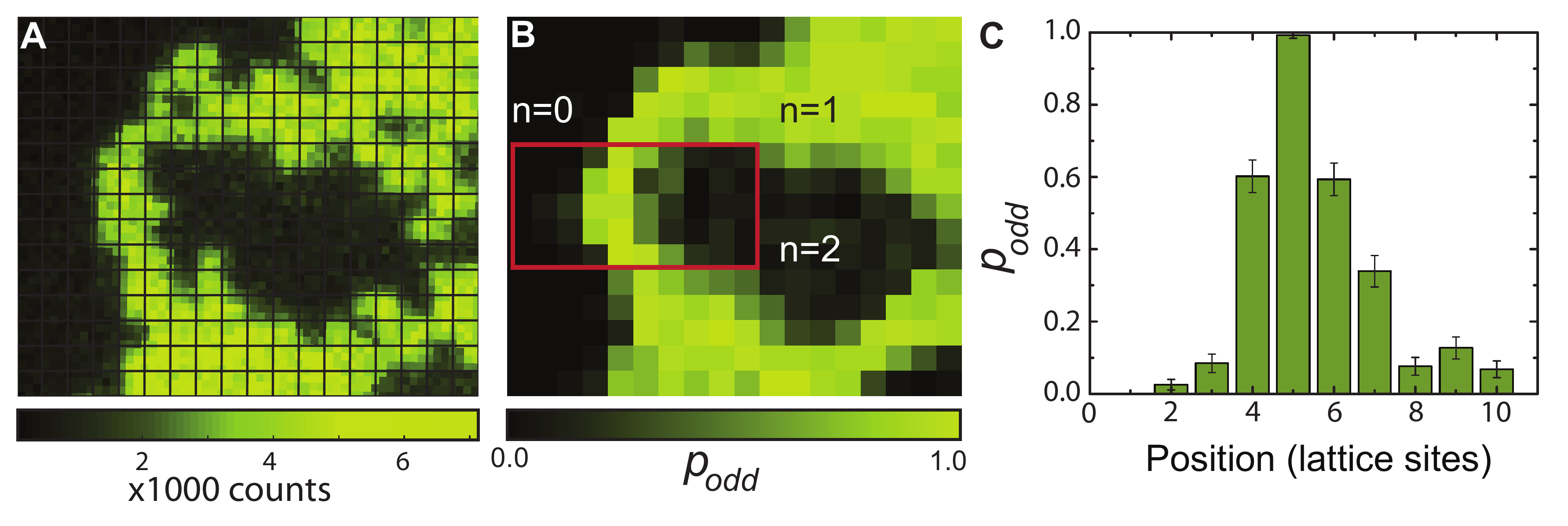}
    \caption{\label{fig4} Low entropy Mott domains observed in a steep
    potential gradient. ({\bf A}) Single shot in-situ image of a Mott
    insulator in a $16E_r$ deep lattice with 25Hz transverse confinement. The
    ring is an $n=1$ insulator enclosing an $n=2$ region. ({\bf B}) Average
    $p_{\mbox{\scriptsize \emph{odd}}}$ over 24 images. Each pixel corresponds
    to a single lattice site.  The red rectangle encloses a region containing
    a Mott insulator with $n=1$, a few lattice sites wide. ({\bf C}) Column
    average of $p_{\mbox{\scriptsize \emph{odd}}}$ over the sites within the
    red rectangle in ({\bf B}), with statistical errorbars.}
\end{figure}

Although the local number statistics change on a fast timescale of $h/U$,
entropy redistribution in the inhomogeneous potential should occur on a much
slower timescale of $h/J$. Because superfluid and normal domains have a larger
specific heat capacity than Mott domains, in an inhomogeneous system, entropy
is expelled from the Mott domains and accumulates in the transition regions
after crossing the phase transition if the system is in thermal equilibrium
\cite{pollet_temperature_2008}. It was found, however, that in bulk Mott
regions the insulating behavior makes entropy transport difficult, and global
thermalization is slow on experimental timescales \cite{hung_slow_2010}. In
our system, optical potential corrugations produce sizable potential gradients
in some regions, leading to a heterostructure of almost one-dimensional Mott
domains, about 1-2 lattice sites thick, surrounded by transition layers (Fig.
\ref{fig4}). We find remarkably low defect densities and sharp transitions between
superfluid and Mott states in these regions. The measured defect probability
per site in the domain shown is $0.8 \pm 0.8\%$. In these microscopic domains,
each site of a Mott domain is in contact with a superfluid region. Such a
configuration is likely to lead to fast thermalization, which would explain
the low defect density we observe. This suggests that the lowest entropies in
a Mott insulator might be obtained under conditions where the chemical
potential is engineered so as to obtain alternating stripes (2D) or layers
(3D) of insulating and superfluid regions
\cite{popp_ground-state_2006,capogrosso-sansone_monte_2008}.

In addition to the number statistics studied in this work, single-site imaging
could be applied to study spatial correlations in strongly correlated quantum
gases\cite{kapit_even-odd_2010}, and to directly measure entanglement in a
quantum information context. The low defect Mott states we detect would
provide an ideal starting point for quantum magnetism experiments; if the low
entropy in the Mott domains can be carried over to spin models, it should be
possible to realize magnetically ordered states such as antiferromagnets,
which could be directly detected with single-site imaging.

\appendix*
\section{Materials and Methods}
\subsection{Preparation of the two-dimensional condensate}
A nearly pure Bose-Einstein condensate of $5\times 10^4$ $^{87}$Rb atoms is
prepared in the $F=1, m_f= -1$ state in a magnetic trap by radiofrequency (rf)
evaporation. The Thomas-Fermi radii of the condensate are (3.1,3.1,27)$\mu$m.
The condensate is transferred into a single well of a 1D standing wave with
periodicity 9.2$\mu$m created by a beam reflected from the flat glass surface of
an in-vacuum hemispheric lens. The light for this standing wave is centered at
755nm, has a 3nm spectral width and is incident at an angle of 2.3$^\circ$
relative to the surface. The condensate is loaded into the first nodal plane
from the surface. The harmonic oscillator width of the condensate at full
lattice depth along the direction perpendicular to the surface is 360nm.
By increasing the bias field the confinement in the 2D plane is relaxed,
resulting in an elliptic cloud with Thomas-Fermi radii (18, 36)$\mu$m in the 2D plane.

In order to obtain a suitable initial density for creating a Mott insulator,
the atom number in the 2D plane must be reduced to a few thousand atoms in a
reproducible way. For this, a red-detuned (840nm) beam with an 8$\mu$m waist
is focused through the objective onto the center of the pancake and creates a
``dimple'' potential in the magnetically confined cloud.  The magnetic
confinement is then removed and the number of atoms remaining in the dimple
trap is proportional to its depth, with a residual RMS fluctuation of 6\%. A
second collinear 840nm beam with a 27$\mu$m waist is then turned on, and the
dimple is adiabatically ramped down to expand the cloud into the larger beam.
The transverse confinement of the condensate provided by this beam is 9.5Hz.
The 840nm light source used for creating these beams has a spectral width of
12nm. The short coherence length eliminates unwanted interferences which would
corrugate the confining potential.

The interaction between the atoms is then further enhanced by increasing the
axial trapping frequency by a factor of eight. This is achieved by turning on
a second 1D standing wave at an angle of 14$^\circ$ to the glass surface, with
a lattice spacing of 1.54$\mu$m. The maximum axial trapping frequency is
7.1kHz. The condensate resides in the sixth well from the surface. At this
point, the 9.2$\mu$m standing wave is ramped down. The surface provides a
reproducible way to overlap the nodes of these two standing waves. In
addition, the proximity of the atoms to the glass surface enhances the
resolution of the imaging system by the index of refraction of glass,
resulting in a measured point spread function (PSF) with full width at half
maximum of $\sim600$nm.

\subsection{Preparation and imaging of the Mott insulator}
To bring the cloud into a strongly-correlated regime, a two-dimensional square
lattice with 680nm periodicity is created in the plane by projecting a mask
through the objective onto the atoms, as described in
\cite{bakr_quantum_2009}. The lattice light, like the light used for producing
the 1D standing waves, is centered at 755nm and has a spectral width of 3nm to
reduce disorder in the potential.

The lattice depth is increased linearly to $0.4E_r$ in 50ms, and from there
ramped exponentially to its final value ($16E_r$ for most experiments) with a
time constant of 81ms. During the lattice ramp, the transverse confinement is
increased so as to keep the size of the cloud constant, compensating for the
increasing inter-atomic interaction and deconfinement due to the blue lattice.
This allows for faster ramps while maintaining adiabaticity, because the
density redistribution during the lattice ramp is minimized. Lattice depths
are calibrated to 5\% accuracy using Kapitza-Dirac scattering, and the
tunneling matrix element and interaction energy at different depths are
obtained from a band structure calculation. The dominating loss process in the
lattice is three-body collisions in Mott shells with $n>2$, as observed in
other experiments \cite{hung_slow_2010}. The rate for such losses is $\gamma
n(n-1)(n-2)$, with $\gamma=2\times10^{-3}$Hz for a lattice depth of $22E_r$
\cite{jack_signatures_2003}. Due to the relatively large lattice spacings,
such losses are negligible in our lattice even for the fourth Mott shell
($<1$\%).

For imaging the atoms, the same procedure as described in
\cite{bakr_quantum_2009} is used. Briefly, the lattice depth is increased over
three hundred-fold by changing the light source illuminating the mask to a
monochromatic source detuned 50GHz to the blue of the D1 line. The frozen atom
distribution, now projected onto number states in each well, is illuminated
with a cooling molasses on the D2 line. Within the first 100$\mu$s, light
assisted collisions eject atoms in pairs, leaving behind an atom only if the
initial atom number on the site was odd. The remaining atoms are imaged in
fluorescence by collecting the scattered molasses photons during a 1s
exposure, resulting in $\sim2,000$ photons registered by the camera for each
atom.

Atom numbers are measured by switching off the transverse confinement and
letting the cloud expand in the 2D plane for 5ms before turning on the deep
lattice used for fluorescence imaging. This ensures that the probability of
two atoms being on the same site is negligible, avoiding photo-assisted losses
for accurate atom number determination.

\subsection{Image analysis}
A sparse atom cloud image is used to extract the PSF. The geometry of the
lattice is then extracted from such an image. First, the lattice spacing is
obtained and then the region of interest is fitted in blocks of 10 by 10
lattice sites. The block centers are allowed to vary to extract any
distortions of the lattice pattern due to imaging aberration over the field of
view. A histogram of atom brightness is used to set a threshold that
identifies the presence or absence of an atom on a site. The information about
the PSF, lattice geometry and threshold obtained from these sparse images is
then used to fit other images with much higher lattice filling, only allowing
for a single global offset in the lattice phase determined by fitting atoms at
the edges of the cloud.

During imaging, a small fraction of the atoms are lost due to background gas
collisions. If this occurs before they scatter enough photons to surpass the
detection threshold, they are not counted. The mean fraction of such uncounted
atoms is $1.75 \pm 0.02\%$, determined from 15 movies (30 frames, 0.5s
exposure per frame) of the atom population decay in the near-resonant lattice.

\subsection{Correction of disorder in optical potentials}
Scatterers on optical surfaces produce ring-like patterns on the optical potentials used to trap the atoms.The spatial pattern of the disorder is static in time and for a lattice depth of $22E_r$, has an RMS gradient of $(0.13\pm0.01)U$ per lattice
site and a characteristic length scale of 10 lattice sites.  The contour lines of the potential are directly extracted from the shell structure boundaries in the Mott regime. Different contour lines are obtained by varying the atom number. The ability to observe these contour lines enables us to correct the potential by projecting an appropriate light pattern through the objective. This pattern is produced by illuminating a digital micromirror device (DLP Discovery 4100, Texas Instruments) with incoherent light of spectral width 1nm, centered at 840nm.  A block of $14 \times 14$ mirrors maps onto a single lattice site in the plane of the atoms, allowing the creation of grayscale patterns, with the aperture of the objective providing Fourier filtering. An error diffusion algorithm \cite{liang_1.5_2009} is used to convert the desired grayscale image to a binary pattern. Potential corrections of either sign are possible by operating the micromirror device with a bias light level produced by flattening the profile of the Gaussian illumination beam.

\begin{acknowledgments}
We would like to thank G. Jotzu, E. Demler, D. Pekker, B. Wunsch, T. Kitagawa,
E. Manousakis, and M. D. Lukin for stimulating discussions. This work was
supported by a grant from the Army Research Office with funding from the DARPA
OLE program, grants from AFOSR MURI, NSF, the Swiss National Science
Foundation, and an Alfred P. Sloan Fellowship to M.G. The simulations were run
on the Brutus cluster at ETH Zurich.
\end{acknowledgments}

\bibliography{mott_paper_arXiv}

\end{document}